\documentclass{emulateapj}
\usepackage{apjfonts}
\shorttitle{\sc Path Density and Quasar Absolute Magnitude}
\shortauthors{\sc Evans \etal}
\input{iondefs.sty}
\begin{document}
\title{The Redshift Distribution of Intervening Weak {\MgII} Quasar
Absorbers and a Curious Dependence on Quasar Luminosity} 
\author{\sc Jessica L. Evans,\altaffilmark{1} Christopher
W. Churchill,\altaffilmark{1} Michael T. Murphy,\altaffilmark{2},
Nikole M. Nielsen,\altaffilmark{1} and Elizabeth
S. Klimek\altaffilmark{1}}
\altaffiltext{1}{New Mexico State University, Las Cruces, NM 88003}
\altaffiltext{2}{Centre for Astrophysics and Supercomputing, Swinburne
University of Technology, Hawthorn, Melbourne, VIC 3122, Australia}
\begin{abstract}
We have identified 469 {\MgIIdblt} doublet systems having $W_r \geq
0.02$ {\AA} in 252 Keck/HIRES and UVES/VLT quasar spectra over the
redshift range $0.1 < z < 2.6$.  Using the largest sample yet of 188
weak {\MgII} systems ($0.02$ {\AA} $\leq W_r < 0.3$ {\AA}), we
calculate their absorber redshift path density, $dN/dz$.  We find
clear evidence of evolution, with $dN/dz$ peaking at $z \sim 1.2$, and
that the product of the absorber number density and cross section
decreases linearly with increasing redshift; weak {\MgII} absorbers
seem to vanish above $z \simeq 2.7$.  If the absorbers are ionized by
the UV background, we estimate number densities of $10^6 - 10^9$ per
Mpc$^3$ for spherical geometries and $10^2 - 10^5$ per Mpc$^3$ for
more sheetlike geometries.  We also find that $dN/dz$ toward
intrinsically faint versus bright quasars differs significantly for
weak and strong ($W_r \geq 1.0$ {\AA}) absorbers.  For weak
absorption, $dN/dz$ toward bright quasars is $\sim 25\%$ higher than
toward faint quasars (10~$\sigma$ at low redshift, $0.4 \leq z \leq
1.4$, and 4~$\sigma$ at high redshift, $1.4 < z \leq 2.34$).  For
strong absorption the trend reverses, with $dN/dz$ toward faint
quasars being $\sim 20\%$ higher than toward bright quasars (also
10~$\sigma$ at low redshift and 4~$\sigma$ at high redshift).  We
explore scenarios in which beam size is proportional to quasar
luminosity and varies with absorber and quasar redshifts.  These do
not explain $dN/dz$'s dependence on quasar luminosity.
\end{abstract}
\section{Introduction}
\label{sec:intro}
Quasar absorption line systems are an extremely useful means of
statistically constraining the various scenarios of metal enrichment,
inflow and outflow, ionization conditions, kinematics, and gas
structure within galaxies and the IGM.  Various {\MgII} absorption
line studies have concluded that these systems are cosmologically
distributed \citep{lanzetta87, sargent88, steidel92}, and numerous
subsequent studies have identified specific galaxies associated with
{\MgII} absorption \citep[see][]{bb91, sdp94, steidel97, guillemin,
cwc-china, chen08, ggk08, barton09, chen10, kacprzak11, churchill12,
nielsen12}.

For the following discussion, we adopt the terms ``weak'',
``intermediate'', and ``strong'' to refer to absorbers having $0.02$
{\AA} $\leq W_r < 0.3$ {\AA}, $0.3$ {\AA} $\leq W_r < 1.0$ {\AA}, and
$W_r \geq 1.0$ {\AA}, respectively, where $W_r$ is the rest frame
equivalent width of the {\MgII} $\lambda$2796 transition.
{\MgII}-selected gas probes a wide range of {\HI} column density
environments.  Weak {\MgII} absorption in particular, which samples
optically thin gas over a large span of cosmic time, has been proposed
to sample dwarf or LSB galaxies as well as the IGM
\citep{weakI,rigby02}.  However, there are instances in which weak
{\MgII} is identified in the circumgalactic medium of ``normal''
bright galaxies \citep{churchill12}.  \citet{milutinovic06}, using an
ionization model, concluded that filamentary and sheetlike IGM
structures host at least a portion of weak {\MgII} absorption.
Moreover, \citet{nielsen12} argue that $W_r < 0.1$ {\AA} absorbers
likely reside in the IGM.  Photoionization modeling of weak {\MgII}
systems with associated {\CIV} led \citet{lynch07} to argue for a
scenario of a shell geometry as might be expected for supernova
remnants or high velocity clouds moving in a hot corona.

Better insight into the nature of the structures selected by weak
{\MgII} absorbers remains elusive and partially motivates this study.
Measurements of the redshift path density $dN/dz$ of weak {\MgII}
absorbers place important constraints on $n(z) \sigma(z)$, the product
of their number density and cross section \citep{weakI, rigby02,
anand07}.  In the same vein, probing discrepancies in redshift path
densities of {\MgII} absorption in various equivalent width ranges
based on differences in the background sources may reveal information
about these absorbing structures.  Despite the evidence indicating
that absorption line systems are primarily associated with intervening
gas as opposed to the background source itself, several studies have
nevertheless revealed major discrepancies, depending on the nature of
the background source, in the incidence of these absorbers per unit
redshift.

\citet{stocke97}, in a study of strong {\MgII} absorbers in BL Lac
objects, observed a redshift path density of 4--5 times greater than
that expected from quasar surveys.  \citet{prochter06b} compared quasar
and GRB sightlines and found the latter to have a factor of $\sim 4$
excess in the redshift path density of strong absorbers.  A similar
study by \citet{vergani09} found a factor of $\sim 2$ excess of strong
absorbers, but no excess of intermediate absorbers, in the GRB sample.
\citet{bergeron11} reported a factor of $\sim 2$ excess of both strong
and intermediate absorbers in blazar versus quasar sightlines. It
should be noted, however, that in that study the authors did find
marginal statistical evidence in the strong sample of an excess
occurring nearer the blazar (even after excluding absorbers having
velocity separations of less than 5,000 {\kms} from the emission
redshift).  Finally, in a study analyzing {\MgII} absorbers down to
$W_r = 0.07$ {\AA} in GRB sightlines, \citet{tejos09} found a factor
of $\sim 3$ overabundance of strong absorbers and a $\sim 30\%$ {\it
reduction} for $0.07 \leq W_r < 1.0$ {\AA} compared to studies of
quasar sightlines.  The latter discrepancy, however, was deemed
insignificant since the results were consistent at the $1~\sigma$
confidence level.  \citet{tejos09} have so far presented the only
study to compare redshift path density differences in weak {\MgII}
absorbers based on different types of background sources.

In this paper we present the largest study of weak {\MgII} absorption
to date, and for the first time provide a parameterized fit to the
redshift path density evolution.  We also present our findings of
differential absorber redshift path density, based on the absolute
magnitude of the background quasar, for weak, intermediate, and strong
{\MgII} systems.  In {\S} \ref{sec:data} we present our data, and in
{\S} \ref{sec:results} our results.  We discuss plausible
interpretations in {\S} \ref{sec:discuss}, and conclude in {\S}
\ref{sec:conclude}.  The cosmological parameters $H_0 = 70$
{\kms}~Mpc$^{-1}$, $\Omega_m = 0.3$, and $\Omega _\Lambda = 0.7$ are
adopted throughout.
\section{Data and Subsamples}
\label{sec:data}
We have searched 252 Keck/HIRES and UVES/VLT quasar spectra for
{\MgIIdblt} doublet absorption.  All systems were objectively
identified using the methods of \citet{schneider93} and \citet{weakI}.
Further details are provided in \citet{evans-phd}\footnote{\it
http://astronomy.nmsu.edu/jlevans/phd} and \citet{evans2012}.  The
search space omitted redshifts within 5,000~{\kms} of the quasar
emission redshift, $z_{em}$, and blueward of the {\Lya} emission of
the quasar.  A total of 422 absorbers comprise our sample.

We divided the absorbers into three subsamples using historically
motivated weak \citep{weakI}, intermediate \citep{steidel92,nestor05},
and strong \citep{steidel92,nestor05} equivalent width ranges.  In
order to determine whether our absorber subsamples are consistent with
being cosmologically distributed along the lines of sight to the
quasars, we performed the second test of \citet{bahcall69}.
Performing the Kolmogorov-Smirnov (KS) test for each of our equivalent
width ranges, we could not rule out that their distributions are
consistent with being cosmological\footnote{We calculated the
\citet{bahcall69} $Y$ parameter of each {\MgII} system, where $0 \leq
Y \leq 1$, with $Y = 0$ representing a doublet at the minimum observed
redshift included in the search, and $Y = 1$ representing a doublet at
the maximum observed redshift.  Following the formalism of
\citet{steidel92}, the sensitivity function $g(Y)$ of the survey was
then calculated.  This is a measure of the number of lines of sight in
the survey in which a system of a given minimum $W_r$ could have been
detected at each value of $Y$.  The $Y$ distributions of each absorber
sample were then statistically compared to $g(Y)$.}.  This result is
in agreement with earlier studies \citep{lanzetta87, sargent88,
steidel92}.

Since none of the quasars in our sample were observed with {\it a
priori\/} knowledge of weak absorption, our survey is unbiased for
this population.  However, since the quasar sample is drawn from a
broad range of targeted science programs, there is the possibility of
bias in the intermediate and strong absorbers, which are known to
sometimes be associated with DLAs \citep{rao00}, or which in some
cases were already known due to previous lower resolution surveys.

To examine whether our intermediate and strong absorber samples are
consistent with an unbiased population, we compared our measured
equivalent width distributions, $f(W_r,W_{\ast})$, where $W_{\ast}$ is
the characteristic $W_r$, to the distribution measured by
\citet{nestor05}.  Using the KS test for the three redshift bins
measured by \citet{nestor05}, we obtained $P({\rm KS}) = 0.053$
($0.36 \leq z \leq 0.87$), $0.621$ ($0.87 \leq z \leq 1.31$) and
$0.322$ ($1.31 \leq z \leq 2.27$), respectively.  For the full
redshift range encompassing all three bins, we obtained $P({\rm KS})
= 0.115$.  Even in the case of the lowest value of $P({\rm KS})$,
corresponding to the lowest redshift range, the two populations are
not inconsistent with each other to even a $2~\sigma$ level.  We thus
proceed under the assumption that our sample of absorbers is a fair
sample.

For our analysis in {\S} \ref{sec:results}, we obtained absolute
$B$-band and apparent magnitudes (primarily $B$, $V$, and $R$) of the
quasars.  The majority were obtained from \citet{vcv01}; 16 were
obtained from the NASA/IPAC Extragalactic Database; and for two
quasars, the magnitudes could not be determined so these lines of
sight and their absorbers were omitted from analysis for which these
quantities were required.
\section{Results}
\label{sec:results}
Following the formalism of \citet{lanzetta87}, modified to account for
the doublet ratio \citep{weakI}, we calculated the number of absorbers
per unit redshift, $dN/dz$.
\subsection{Weak Absorber Redshift Path Density}
\label{sec:weakdndz}
For our full redshift range, $0.1 \leq z_{abs} \leq 2.6$, the
cumulative redshift path is $\Delta Z = 231$ and is $\sim100$\%
complete to a 5 $\sigma$ equivalent width sensitivity of $W_r(2796) =
0.05$ {\AA} and $\sim82$\% complete to a 5 $\sigma$ equivalent width
sensitivity of $W_r(2796) = 0.02$ {\AA}.  Over the redshift range $0.4
\leq z_{abs} \leq 2.4$, the extent of a study by \citet{anand07}, we
have $\Delta Z = 213$, compared to their 70; and for $0.4 \leq z_{abs}
\leq 1.4$, the extent of a study by \citet{weakI}, we have $\Delta Z =
148$, compared to their 17.  These represent the largest two previous
weak {\MgII} surveys.  Our larger cumulative redshift path reflects
the larger number of lines of sight included in our survey.
\citet{weakI} surveyed 26 HIRES quasar spectra and found 30 weak
systems, while \citet{anand07} surveyed 81 UVES quasar spectra and
found 112 weak systems.  In our survey we identified 188 weak systems.

We calculated $dN/dz$ for weak systems in four redshift bins in order
to facilitate comparison with the works of \citet{anand07} and
\citet{weakI}; the bins and results are shown in Table
\ref{tab:dNdzcomp}.
\begin{deluxetable}{ccccc}
\tabletypesize{\scriptsize}
\tablewidth{0pt}
\tablecaption{$dN/dz$~ for {\MgII} Surveys of $0.02$ {\AA} $\leq W_r < 0.3$ {\AA}\label{tab:dNdzcomp}}
\tablehead{
\colhead{Survey} & 
\colhead{$0.4\!<\!z\!<\!0.7$} & \colhead{$0.7\!<\!z\!<\!1.0$} & \colhead{$1.0\!<\!z\!<\!1.4$} & 
\colhead{$1.4\!<\!z\!<\!2.4$}
}
\startdata
CRCV99\tablenotemark{a} & $1.43 \pm 0.21$ & $1.84 \pm 0.26$ & $2.19 \pm 0.80$ & $\cdots$       \\[2pt]
NMCK07\tablenotemark{b} & $1.06 \pm 0.10$ & $1.51 \pm 0.09$ & $1.76 \pm 0.08$ & $1.06 \pm 0.04$\\[2pt]
this survey             & $0.74 \pm 0.02$ & $1.08 \pm 0.02$ & $0.95 \pm 0.02$ & $0.67 \pm 0.01$\\[-5pt]
\enddata
\tablenotetext{a}{\citet{weakI}}
\tablenotetext{b}{\citet{anand07}}
\end{deluxetable}
%

All three studies obtained different results, with the trend being
that the $dN/dz$ values have decreased with larger survey size.  All
three used the identical code \citep[SEARCH,][]{weakI} for line and
candidate identification, and all three ostensibly used the same
algorithms in determining the redshift paths for each system.  To test
for possible differences in the calculations\footnote{The
\citet{weakI} spectra were a subset of the quasars we searched.}, we
ran our code used for this survey on the identical spectra and set of
systems used by \citet{weakI} and compared their $\Delta Z$ and
$dN/dz$ results with those of the reproduced study.  The result was
that the $\Delta Z$ values from the \citet{weakI} study were $\sim
30\%$ larger than our reproduced values from their data, and the
$dN/dz$ values from the original study were thereby lowered compared
to our reproduced study.  This suggests that our redshift path
calculations are more conservative than those of \citet{weakI}.  If
our redshift paths had been calculated exactly as theirs, the $dN/dz$
result of this survey would presumably have been lower, further
widening the discrepancy with previous works.  A similar duplication
of the survey of \citet{anand07} could not be performed because a
significant number of the authors' quasar spectra were unavailable to
us.

In an attempt to find differences in the quasar samples that could
possibly lead to the discrepant $dN/dz$ results among the three works,
we investigated the quasar apparent magnitude distributions.  The KS
test was performed between all possible pairs of the three surveys.
Our survey and that of \citet{anand07} exhibit remarkably similar
distributions; it could not be ruled out to greater than a $1 \sigma$
confidence level that their apparent magnitudes had been drawn from
the same population, and their median values were both 17.5.  In
contrast, the \citet{weakI} survey differed from each of the other two
to a confidence level of $4 \sigma$ and had a median apparent
magnitude of 16.3.  A uniform set of apparent magnitudes in the same
band was not available, however, making these comparisons uncertain.
The smaller \citet{weakI} survey was undoubtedly overall biased toward
brighter quasars.

We likewise investigated the quasar {\it absolute} $B$-band magnitude
distributions, but when each survey was tested against the other two,
it could not be ruled out to even a $1 \sigma$ confidence level that
their distributions had been drawn from the same population.
Similarly, no substantial differences were found in the overall median
quasar absolute magnitude of the three surveys: $-28.5$, $-28.7$, and
$-28.6$ for \citet{weakI}, \citet{anand07}, and this study,
respectively.

The reasons behind the different $dN/dz$ results among the three works
remain unclear; future studies will hopefully resolve the weak {\MgII}
puzzle.  It may be that the inclusion or exclusion of weak systems
very close to the limiting equivalent width may play a role (Anand
Narayanan, private communication), as well as differences in the codes
used, since the calculation of $\Delta Z$ is extremely sensitive for
the weakest systems.
\subsection{Redshift Evolution}
\label{sec:evolution}
\begin{figure}[htb]
\figurenum{1} \epsscale{1.2} \plotone{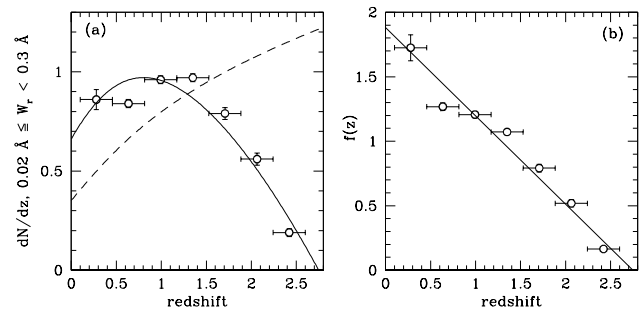}
\caption{
($a$) shows the weak $dN/dz$ (open circles) as well as the NEE
(dashed line).  --- ($b$) shows the same data divided by the NEE.  The
solid line is the best linear fit to this ratio.  In ($a$), this
fit was then multiplied by the NEE and plotted (solid curve).}
\label{fig:residual}
\end{figure}

The number of absorbers per unit redshift is the product of the proper
number density of absorbers and their proper geometric cross section.
In the standard cosmological model, the no-evolution expectation (NEE)
for the redshift number density can be parameterized as:
\begin{equation}
\label{equationdndzparameterization}
\left[{dN \over dz}\right]_{\hbox{\tiny NEE}} = {c\over
H_{0}}n_0\sigma_0{(1+z)^{2}\over\sqrt{\Omega_{m}(1+z)^{3}+\Omega_{\Lambda}}}
\end{equation}
where $n_0$ is the mean comoving number density of absorbers and
$\sigma_0$ is the mean comoving geometric absorber cross section.  As
shown in Figure~\ref{fig:residual}$a$, our observed $dN/dz$ departs
strikingly from the NEE (dashed curve), which was normalized to the
mean $z_{abs}$, 1.098, and to the overall $dN/dz$, 0.83, of our weak
sample over the redshift range $0.1 \leq z_{abs} \leq 2.6$. The
general behavior of $dN/dz$ is in agreement with \citet{anand07} in
that it peaks between $1.0 < z < 1.4$ and then decreases toward higher
redshift.

If the product $n \sigma$ varies as a function of redshift, $dN/dz$ may
depart from the no-evolution expectation.  The quantity $n(z)\sigma
(z)$ can be written as
\begin{equation}
n(z)\sigma (z) = n_0\sigma_0f(z),
\end{equation}
where $f(z)$ is a nonnegative function that parameterizes the
evolution of $dN/dz$.

Figure~\ref{fig:residual}$b$ depicts our weak $dN/dz$ result
divided by the NEE.  The data clearly motivate a linear fit; this was
achieved using a function of the form
\begin{equation}
f(z) = 1 - \alpha(z-z^*),
\end{equation}
where $\alpha$ is the slope and $z^*$ is the function normalization.
The result, $\alpha = 0.69\pm0.02$ and $z^* = 1.29\pm0.05$, is shown
as a solid line in Figure~\ref{fig:residual}$b$.  

Finally, the fit to $f(z)$ was multiplied by the NEE, yielding a fit
to the $dN/dz$ result in Figure~\ref{fig:residual}$a$ (solid curve)
that can be expressed as
\begin{equation}
\label{equationfit}
{dN \over dz} = {c\over H_{0}}n_0\sigma_0{(1+z)^{2}\left[1 -
\alpha(z-z^*)\right]\over
\sqrt{\Omega_{m}(1+z)^{3}+\Omega_{\Lambda}}} \qquad (z<2.74) \, .
\end{equation}
Weak {\MgII} absorbers seem to vanish at high redshift; this result is
discussed in {\S} \ref{sec:discuss}.
\subsection{Differential Absorber Redshift Path Densities by Absolute Magnitude}
\label{sec:mag}
Motivated by studies that found differences in $dN/dz$ results based
on background object as discussed in {\S} \ref{sec:intro}, as well as
our finding of differing apparent magnitude distributions among the
weak {\MgII} studies as discussed in {\S} \ref{sec:weakdndz}, we
attempted to discern some intrinsic difference that might affect the
observed $dN/dz$.  Using the absolute $B$-band quasar magnitudes of
our survey, which had a range of $-32.2 \leq M_B \leq -19.8$, we
divided our quasars into ``bright'' and ``faint'' subsamples according
to the median, $\left< M_B \right> = -28.6$.  The bright subsample has
a median of $-29.4$, and the faint subsample has a median of $-27.4$.
\begin{deluxetable}{lccccccccc}
\tabletypesize{\tiny}
\tablewidth{0pt}
\tablecaption{$dN/dz$ by Quasar Luminosity\label{tab:dNdz}}
\tablehead{
\colhead{\tiny sample}&
\colhead{\tiny weak}&\colhead{\tiny sig}&
\colhead{\tiny intermediate}&\colhead{\tiny sig}&
\colhead{\tiny strong}&\colhead{\tiny sig}\\
\colhead{\tiny \phantom{x}}&
\colhead{\tiny $0.02 \leq W_r < 0.3$}&\colhead{\tiny lev}&
\colhead{\tiny $0.3 \leq W_r < 1.0$}&\colhead{\tiny lev}&
\colhead{\tiny $W_r \geq 1.0$}&\colhead{\tiny lev}
}
\startdata
\cutinhead{\tiny $0.4 \leq z_{abs} \leq 1.4$}
{\tiny all}       &{\tiny $0.931 \pm 0.006$}&$$                      &{\tiny $0.715 \pm 0.005$}&$$                      &{\tiny $0.449 \pm 0.003$}&$$\\[0pt]
{\tiny bright}    &{\tiny $1.040 \pm 0.016$}&$$                      &{\tiny $0.692 \pm 0.010$}&$$                      &{\tiny $0.406 \pm 0.006$}&$$\\[0pt]
{\tiny faint}       &{\tiny $0.836 \pm 0.011$}&$$                      &{\tiny $0.730 \pm 0.009$}&$$                      &{\tiny $0.481 \pm 0.006$}&$$\\[0pt]
{\tiny bright/all}&{\tiny $1.117 \pm 0.019$}&{\tiny $6.2~\sigma$}&{\tiny $0.967 \pm 0.016$}&{\tiny $2.1~\sigma$}&{\tiny $0.905 \pm 0.015$}&{\tiny $6.3~\sigma$}\\[0pt]
{\tiny faint/all}   &{\tiny $0.897 \pm 0.013$}&{\tiny $7.9~\sigma$}&{\tiny $1.021 \pm 0.014$}&{\tiny $1.5~\sigma$}&{\tiny $1.072 \pm 0.015$}&{\tiny $4.8~\sigma$}\\[0pt]
{\tiny faint/bright}&{\tiny $0.803 \pm 0.020$}&{\tiny $9.9~\sigma$}&{\tiny $1.056 \pm 0.019$}&{\tiny $2.9~\sigma$}&{\tiny $1.185 \pm 0.019$}&{\tiny $9.7~\sigma$}\\[-5pt]
\cutinhead{\tiny $1.4 < z_{abs} \leq 2.34$}
{\tiny all}       &{\tiny $0.686 \pm 0.011$}&$$                      &{\tiny $0.696 \pm 0.011$}&$$                      &{\tiny $0.631 \pm 0.010$}&$$\\[0pt]
{\tiny bright}    &{\tiny $0.732 \pm 0.018$}&$$                      &{\tiny $0.730 \pm 0.018$}&$$                      &{\tiny $0.584 \pm 0.014$}&$$\\[0pt]
{\tiny faint}       &{\tiny $0.579 \pm 0.028$}&$$                      &{\tiny $0.619 \pm 0.030$}&$$                      &{\tiny $0.714 \pm 0.034$}&$$\\[0pt]
{\tiny bright/all}&{\tiny $1.068 \pm 0.031$}&{\tiny $2.2~\sigma$}&{\tiny $1.050 \pm 0.031$}&{\tiny $1.6~\sigma$}&{\tiny $0.926 \pm 0.027$}&{\tiny $2.7~\sigma$}\\[0pt]
{\tiny faint/all}   &{\tiny $0.845 \pm 0.043$}&{\tiny $3.6~\sigma$}&{\tiny $0.890 \pm 0.045$}&{\tiny $2.4~\sigma$}&{\tiny $1.131 \pm 0.057$}&{\tiny $2.3~\sigma$}\\[0pt]
{\tiny faint/bright}&{\tiny $0.791 \pm 0.054$}&{\tiny $3.9~\sigma$}&{\tiny $0.848 \pm 0.054$}&{\tiny $2.8~\sigma$}&{\tiny $1.222 \pm 0.053$}&{\tiny $4.1~\sigma$}\\[-5pt]
\enddata
\end{deluxetable}
\begin{figure}[htb]
\figurenum{2} \epsscale{1.2} \plotone{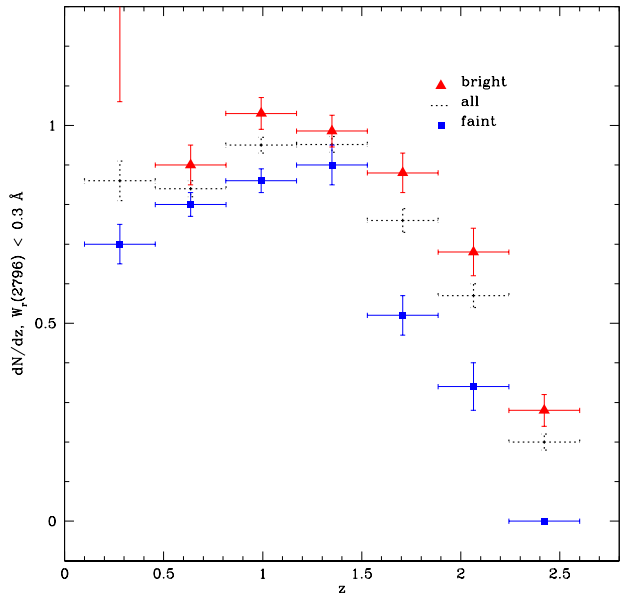}
\caption{
$dN/dz$ of the weak sample for the bright (red triangles), faint (blue
squares), and combined (dotted crosses) samples.  In the bright
sample, the lowest redshift data point has a value of $1.53 \pm
0.47$.}
\label{fig:dndzweakmags}
\end{figure}
\begin{figure}[htb]
\figurenum{3} \epsscale{1.2} \plotone{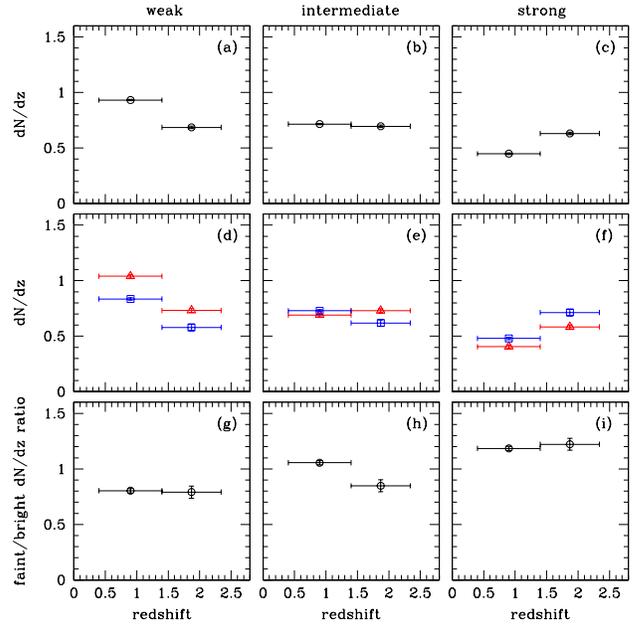}
\caption{
($a$--$c$) show $dN/dz$ of the weak, intermediate and strong samples,
respectively, in two redshift bins ($0.4 \leq z_{abs} \leq 1.4$ and
$1.4 \leq z_{abs} \leq 2.34$).  --- ($d$--$f$) show $dN/dz$ for the
bright (red triangles) and faint (blue squares) subsamples.  The
points were left open to better display the error bars. --- ($g$--$i$)
show the faint to bright $dN/dz$ ratio.}
\label{fig:dndz}
\end{figure}

Figure \ref{fig:dndzweakmags} shows our weak $dN/dz$ results for the
bright and faint samples binned as in Figure \ref{fig:residual}.  The
redshift path density of the bright sample is clearly higher than that
of the faint sample at all redshifts.  The lowest redshift bin of the
weak bright sample contained only five systems and we believe this
caused the $dN/dz$ result in that bin to be less reliable.

Figures~\ref{fig:dndz}$a$--\ref{fig:dndz}$c$ plot our $dN/dz$ results
for weak, intermediate and strong {\MgII} absorption.  The bins ($0.4
\leq z_{abs} \leq 1.4$ and $1.4 \leq z_{abs} \leq 2.34$, ``low'' and
``high'' redshift) were selected based on previous studies
\citep{anand07,nestor05,weakI}.  Similarly,
Figures~\ref{fig:dndz}$d$--\ref{fig:dndz}$f$ show $dN/dz$ calculated
for the faint and bright quasar samples.
Figures~\ref{fig:dndz}$g$--\ref{fig:dndz}$i$ show the ratio of the
$dN/dz$ results of the faint quasar sample to that of the bright.
These values are listed in Table \ref{tab:dNdz} for each $W_r$ and
redshift range.  Though the weak $dN/dz$ results of \citet{weakI},
\citet{anand07}, and this study all differed as discussed in {\S}
\ref{sec:weakdndz}, the {\it relative\/} values among our own faint,
bright and all quasar subsamples are robust, having been calculated in
a consistent manner.  Similarly, though our intermediate and strong
samples have $W_r$ distributions consistent with \citet{nestor05} as
mentioned in {\S} \ref{sec:data}, our $dN/dz$ results for these
samples are higher than those of \citet{nestor05} and
\citet{lundgren09}. This is expected since some of the quasars in our
survey were targeted for their known $W_r \geq 0.3$ {\AA} {\MgII}
absorption; however, the $dN/dz$ ratios between our magnitude bins are
robust.

For weak systems, the $dN/dz$ values toward bright quasars are $\sim
25\%$ higher than for the faint in both redshift ranges, while for
strong systems, $dN/dz$ is $\sim 20\%$ higher toward faint quasars than
toward bright in both redshift ranges.  The faint to bright $dN/dz$
ratio, $(dN/dz)_f/(dN/dz)_b$, departs for the weak systems from unity
at the $\simeq 10~\sigma$ level for low redshift and at the $\simeq
4~\sigma$ level for high redshift.  The strong absorber ratios are
similarly significant (see Table \ref{tab:dNdz}), though in that case
the faint quasar $dN/dz$ values are higher, rather than lower, than
those of the bright sample.  In the case of the intermediate absorbers
the ratio is consistent with unity within $3~\sigma$.

In the calculation of $dN/dz$ the sensitivity of each spectrum is
accounted for \citep[see][]{lanzetta87, evans-phd}, eliminating the
possibility that the higher values of the weak bright sample compared
to the weak faint sample might result from higher signal-to-noise
ratios.

\section{Discussion}
\label{sec:discuss}
Based on our weak $dN/dz$ result (see Figure~\ref{fig:residual}), the
cosmic number density, geometric cross section, or both, of weak
{\MgII} absorbers appear to be evolving.  The apparent dropoff in our
fit toward $z = 0$ may be overly steep; it is possible that the weak
$dN/dz$ peaks at $z \sim 1$, declines slightly and then levels off
toward the present.  However, it is a first attempt to characterize
weak {\MgII} absorber evolution using a functional form.  Our result
predicts that no such absorbers exist above $z \simeq 2.7$.

\citet{anand07} speculated that the apparent paucity of weak {\MgII}
above $z \sim 2$ might be due to the high redshift analogs of low
redshift weak {\MgII} absorption being associated with strong {\MgII}.
In this scenario, weak {\MgII} absorption at high redshift would be in
the kinematic vicinity of strong {\MgII} and thus would not be
recognized as isolated weak absorption.  However, we have compared the
high velocity weak kinematic subsystems of strong {\MgII} subsystems
to isolated weak {\MgII} absorption \citep{evans-phd}.
Morphologically these two types of profiles often appear very similar,
but a KS test of their rest equivalent width distributions revealed
that they are actually two distinct populations to a 99.98\%
confidence level, or greater than 3 $\sigma$.  A KS test of the
distributions of flux decrement-weighted velocity spreads, $\omega_v$
\citep{churchill01, evans-phd} indicated to a greater than 6 $\sigma$
confidence level that the two populations are unique.

The $dN/dz$ evolution we detect in weak {\MgII} may be due to changes
in gas structure or ionization conditions; neither we nor other
studies find the same falloff in intermediate and strong {\MgII}
absorption \citep{nestor05, prochter06a, lundgren09} up to our maximum
redshift of $2.6$.  \citet{matejek12} do report a decline in the
strong population above $z \sim 3$, and note that this peak
corresponds to that of the SFR.  Our weak $dN/dz$ result, which
exhibits no such peak, may provide indirect evidence that a
substantial fraction of these absorbers resides in the IGM, since
their evolution appears not to correspond to star formation.

Using Cloudy 08.00 photoionization modeling \citep{ferland98}, we
investigated the evolution of weak {\MgII} absorber sizes $R(z)$ and
cosmic number densities $n(z)$ \citep[for additional details
see][]{evans-phd}.  In this scenario {\MgII} selects relatively dense
cloudlets embedded within plane parallel slabs of gas.  We modeled
optically thin clouds having a range of hydrogen number densities
based on past {\MgII} photoionization modeling results
\citep{rigby02,bergeron02}.  We assumed an ultraviolet background
model that varies as a function of $z$, following the work of
\citet{haardt96}, which includes the contribution of galaxies.  Though
we examined a grid of clouds with a range of {\HI} column densities
and metallicities, we discuss here clouds having $N(\HI)$ of $10^{16}$
cm$^{-2}$ and a metallicity of 0.1 solar.  For weak {\MgII}, $\log
N({\HI})$ is constrained to the range 15.5--17.0 cm$^{-2}$
\citep{churchill00, rigby02}.

\begin{figure}[htb]
\figurenum{4} \epsscale{1.2} \plotone{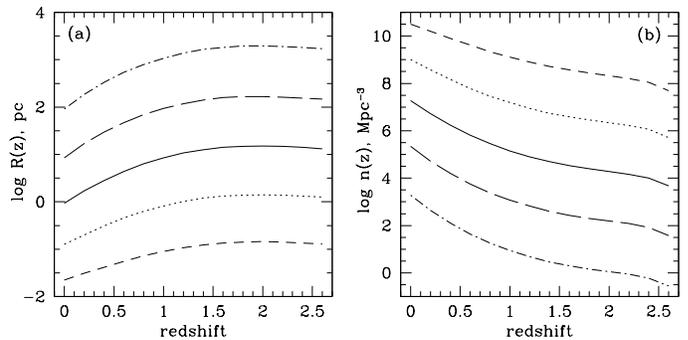}
\caption{
($a$) plots the evolution of Cloudy absorber size, log $R(z)$.  ($b$)
plots the number density of absorbers, log $n(z)$.  The dashed curves
represent log $n_H = -0.5$ cm$^{-3}$; the dotted curves, log $n_H =
-1.0$ cm$^{-3}$; the solid curves, log $n_H = -1.5$ cm$^{-3}$; the
long dashed curves, log $n_H = -2.0$ cm$^{-3}$; and the dashed-dotted
curves, log $n_H = -2.5$ cm$^{-3}$.}
\label{fig:cloudy}
\end{figure}

The resulting cloud thicknesses, which we interpreted as absorber sizes
$R(z)$ and which are governed by the ionizing background, peak at $z
\sim 2$ and then decline toward the present, as shown in
Figure~\ref{fig:cloudy}$a$.  Assuming spherical clouds, the
corresponding absorber cross sections $\sigma (z)$, combined with our
weak $dN/dz$ constraint using the fit of Equation \ref{equationfit},
translate into cosmic absorber number densities $n(z)$ that increase
monotonically toward the present (Figure~\ref{fig:cloudy}$b$).  The
absorber sizes produced by this model are on the order of a parsec,
and yield absorber number densities on the order of $10^6 - 10^9$
Mpc$^{-3}$, for the middle range of $n_H$ values.  This corresponds to
$10^9$--$10^{12}$ absorbers per $L^*$ galaxy for $z \lesssim 1$
\citep{faber07} as well as for $1 \lesssim z \lesssim 3$
\citep{reddy09, oesch10}.

If the clouds are not spherical, but instead the transverse extent
$R_T$ scales with cloud thickness according to a factor $\beta$ such that
$R_T = \beta R(z)$, then $n(z)$ would scale as $\beta ^{-2}$.  For
$\beta = 100$, $n(z)$ would then be reduced by a factor of $10^4$,
which yields $10^2 - 10^5$ weak absorbers per Mpc$^3$.  It should
be noted that changing the model's $N({\HI})$ would change $R(z)$ in
direct proportion, while $n(z)$, using our $dN/dz$ constraints, would
vary as $N({\HI})^{-2}$.

Our Cloudy model is suggestive of a condensation mechanism into sheet
or filament structures characteristic of the IGM.  Sheetlike
geometries require far fewer weak absorbers than do spherical
geometries per $L^*$ galaxy, and therefore we consider these to be a
more realistic scenario.  Although we do not fully explain the nature
of weak {\MgII} absorption, this exercise does provide a limiting case
in which to couch the phenomenon.  Our absorber size estimate is not
far from that of \citet{rigby02}, who concluded that a sheetlike
structure containing many embedded $\sim10$ pc absorbers was required
to account for the observed $dN/dz$.  Recent findings by
\citet{churchill12} and \citet{nielsen12} indicate that $0.1 \leq W_r
\leq 0.3$ {\AA}, but not $0.02 \leq W_r \leq 0.1$ {\AA}, absorbers are
found in the circumgalactic medium of normal galaxies at impact
parameters of less than 200 kpc.  They suggest that the weaker of
these two populations may then reside primarily in the IGM.

Our $dN/dz$ results toward faint versus bright absolute magnitude
quasars reveal that the weak absorbers have a higher redshift path
density in the bright quasar sample than in the faint at both low and
high redshifts.  For the strong absorbers, the opposite is true.  The
intermediate absorbers appear to follow no clear trend, and may
represent an equivalent width range where the effects leading to the
weak and strong differentials mostly cancel.

Following the discovery by \citet{prochter06b} of the quasar--GRB strong
{\MgII} discrepancy, various researchers have attempted to explain the
phenomenon.  \citet{frank07} modeled the effects of {\MgII} absorber
size and impact parameter on observed equivalent width and concluded
that the $dN/dz$ discrepancy may be due to the larger beam sizes of
quasars versus GRBs, the latter of which they state are on the order
of the sizes of cloud cores.  The authors also predicted that
different luminosity populations of quasars should contain different
incidences and strengths of intervening absorbers.

\citet{porciani07} countered this differential beam size argument,
noting that no unsaturated {\MgII} doublets have been observed having
a doublet ratio of one, as would be expected in the case of partial
covering of a quasar beam by an absorber.  They stated that
magnification bias could explain the discrepancy, and that dust
obscuration bias and association of absorbers with the circumburst
environment could also partially account for it.

A statistical study was also conducted by \citet{pontzen07} to look
for systematically lowered {\MgII} equivalent widths over quasar broad
line emission regions, which are substantially larger than quasar
continuum regions; no significant difference was found.

\citet{cucchiara09} cite an intrinsic origin as a possible explanation
for the GRB excess.  {\MgII} absorbing gas could be ejected at
relativistic velocities and masquerade as an intervening absorber;
however, the authors note that the presence of {\MgI} absorption in
these systems, as well as the lack of fine structure transitions that
are expected in the vicinity of a GRB, cast doubt on this theory.
\citet{vergani09} concur that the excess could be intrinsic, and
estimate required ejection velocities of $10,000 - 25,000$ {\kms}.
They also consider gravitational lensing to be a viable mechanism to
account for the discrepancy.

In two studies of {\CIV} absorbers toward GRBs, \citet{sudilovsky07}
and \citet{tejos07} reported no excess incidence over quasar
sightlines.  \citet{sudilovsky07} speculated that the difference in
the cases of {\MgII} versus {\CIV} absorbers arose partially because
the former introduced more dust extinction than the latter.  However,
dust extinction in {\MgII} absorbers was subsequently modeled
\citep{sudilovsky09}, and the authors concluded that the effect could
only account for $\sim 10\%$ of the quasar--GRB discrepancy.

\citet{tejos09} rejected an intrinsic origin for excess GRB {\MgII}
absorbers due to the lack of both excess {\CIV} absorption and excess
weak and intermediate {\MgII} absorption, and instead favored
gravitational lensing as the relevant mechanism.  \citet{wyithe11}
modeled gravitational lensing in quasar and GRB sightlines and
concluded that it was a feasible explanation for the excess, but that
further GRB data were necessary to support or refute their findings.
They noted that afterglows in which strong {\MgII} systems are found
are brighter than average, implying a greater lensing rate.

Through their modeling of extinction curves toward quasars and GRBs,
\citet{budzynski11} determined that $dN/dz$ toward quasars would be
significantly higher if corrected for dust, and that the correction
varies with redshift.  The discrepancy compared to GRBs arises, the
authors state, because their absorber redshift distribution is shifted
higher than that of quasars, resulting in less loss of detected
absorbers.  They calculated that this effect could account for a
factor of two excess in the GRB $dN/dz$.  

Keeping these previous studies of the quasar--GRB discrepancy in mind,
and in an attempt to understand the possibly related phenomenon we
have uncovered, we attempted to find some other metric within our
faint and bright quasar samples that would shed light on these issues.
The redshift path density depends on the integrated equivalent width
distribution,
\begin{equation}
{dN\over dz} \propto {\displaystyle \int _{W_{min}}^{W_{max}} \!\!\!
f(W,W_*)\, dW} \, .
\label{eq:dndzrat}
\end{equation}
We therefore compared equivalent width distributions for our faint and
bright quasar samples for the various $W_r$ ranges as well as for low,
high, and all redshifts.

We also studied equivalent width distributions binned by relative beam
sizes using $M_B$ as a proxy, i.e. assuming that the square of the
source radius $R_s$ is proportional to the $B$-band luminosity of the
quasar.  The ratio of the source radius for quasar $i$ relative to the
median radius for the full quasar sample can then be written
\citep{shakura73} as
\begin{equation}
\label{equationradii}
\frac{R_{s,i}}{\left< R_s \right>} = 10^{-0.2(M_i-\left< M_B \right>)}.
\end{equation}
We then studied the effect of changing beam size with redshift due to
cosmology.  We calculated the ratio of the relative beam size of
quasar $i$ at the redshift of absorber $j$ to the cross section of the
source:
\begin{equation}
{\sigma_b(z_j)\over \sigma_{s,i}} = \left[ \frac{D_{\hbox{\tiny
A}}(z_j)}{D_{\hbox{\tiny A}}(z_{s,i})} \right]^2
\end{equation}
where $\sigma_{s,i} = \pi R_{s,i}^2$, $D_A(z_j)$ is the angular
diameter distance at the absorption redshift of system $j$, and
$D_A(z_{s,i})$ is the angular diameter distance at the source redshift
of quasar $i$.  Finally, we examined the combined effect of source
size and cosmology, by using Equation \ref{equationradii} to scale
$\sigma_b$.

Using the KS test, none of these equivalent width distributions
yielded significant differences (of at least 3 $\sigma$) between the
faint and bright quasar samples.  Since the $dN/dz$ discrepancy is in
this case of a smaller magnitude than in the case of the quasar--GRB
phenomenon, it may require a larger data set to discern the reasons
behind the observations.

The findings of \citet{budzynski11} do offer an intriguing possibility
by relating absorption redshift distributions to $dN/dz$.  Our faint
absorber samples do have lower median $z_{abs}$ values than our bright
samples across all three equivalent width ranges, probably a result of
the correlation whereby intrinsically more luminous quasars tend to be
selected at higher redshifts.  This result only supports the authors'
dust argument in the case of our weak absorbers, the only sample in
which the absorption incidence is significantly higher toward bright
quasars than faint.  For our weak sample, the median absorption redshift
is $0.87$ in the faint subsample and $1.17$ in the bright.  KS testing,
however, revealed that it could not be ruled out to a greater than
98.32\% confidence level that the faint and bright subsamples are drawn
from the same underlying $z_{abs}$ distribution.

Though several authors have argued for an intrinsic origin for excess
strong {\MgII} absorption toward GRBs versus quasars, this does not
appear to explain the $dN/dz$ discrepancy in the case of our bright
and faint quasar populations.  Our \citet{bahcall69} testing revealed
absorber distributions consistent with cosmological within all
equivalent width and quasar absolute magnitude subsamples, as well as
in the aggregate populations.  The velocities of {\MgII}-selected gas
ejected from a quasar would have to reach large fractions of the speed
of light in order to pass for intervening systems.  It therefore seems
highly unlikely that significant intrinsic absorption could be present
in our sample.
\section{Conclusion}
\label{sec:conclude}
We have found in a survey of 252 quasar spectra that the incidence of
weak {\MgII} absorption evolves markedly, that it peaks at $z \sim
1.2$, and that it is fit by a function that is a product of the
no-evolution expectation with a linear function.  Our linear fit to
the ratio of our $dN/dz$ data to the NEE resulted in a slope of
$\alpha = 0.69\pm0.02$ and a normalization of $z^* = 1.29\pm0.05$ for
the function $f(z) = 1 - \alpha(z-z^*)$.  Our $dN/dz$ result predicts
that no weak {\MgII} absorbers exist above $z \simeq 2.7$.

We find that when our quasar survey is segregated by absolute
magnitude, weak {\MgII} $dN/dz$ is significantly lower in the faint
subsample than in the bright, with faint to bright $dN/dz$ ratios of
$0.80 \pm 0.02$ at low redshift and $0.79 \pm 0.05$ at high redshift.
In contrast, strong {\MgII} $dN/dz$ is significantly {\it higher} in
the faint subsample than in the bright, with faint to bright $dN/dz$
ratios of $1.19 \pm 0.02$ at low redshift and $1.22 \pm 0.05$ at high
redshift.  Intermediate equivalent width absorbers exhibited $dN/dz$
ratios consistent with unity within 3 $\sigma$.  At this time it is
uncertain whether these results stem from some intrinsic property of
the quasars, from some difference in the intervening sightlines, or
from some combination of these factors.  

\acknowledgments We thank Wallace Sargent, Michael Rauch, Jason
Prochaska, and Charles Steidel for their contribution of spectra, and
Anand Narayanan for helpful communications regarding \citet{anand07}.
This research made use of the NASA/IPAC Extragalactic Database (NED),
which is operated by the Jet Propulsion Laboratory, California
Institute of Technology, under contract with NASA.  We are grateful
for NSF grant AST 0708210, the primary funding for this work; JLE was
also supported by a three-year Aerospace Cluster Fellowship
administered by the Vice Provost of Research at New Mexico State
University and by a two-year New Mexico Space Grant Graduate Research
Fellowship.  MTM thanks the Australian Research Council for a QEII
Research Fellowship (DP0877998).

\end{document}